\documentstyle[12pt,a4]{article}
\input{epsf}
\global\arraycolsep=2pt 
\begin{document}
\makeatletter
\def\fmslash{\@ifnextchar[{\fmsl@sh}{\fmsl@sh[0mu]}}
\def\fmsl@sh[#1]#2{%
  \mathchoice
    {\@fmsl@sh\displaystyle{#1}{#2}}%
    {\@fmsl@sh\textstyle{#1}{#2}}%
    {\@fmsl@sh\scriptstyle{#1}{#2}}%
    {\@fmsl@sh\scriptscriptstyle{#1}{#2}}}
\def\@fmsl@sh#1#2#3{\m@th\ooalign{$\hfil#1\mkern#2/\hfil$\crcr$#1#3$}}
\makeatother
%
\thispagestyle{empty}
\begin{titlepage}

\begin{flushright}
TTP 99--19 \\
hep-ph/9904475\\
28.4.99
\end{flushright}

\vspace{0.3cm}
\boldmath
\begin{center}
\Large\bf  Comparing $B \to X_u \ell \bar{\nu}_\ell$ 
           to $B \to X_s \gamma$ \\ and the Determination of 
           $|V_{ub}| / |V_{ts}|$ 
\end{center}
\unboldmath
\vspace{0.8cm}

\begin{center}
{\large Thomas Mannel} and {\large Stefan Recksiegel} \\
{\sl Institut f\"{u}r Theoretische Teilchenphysik,
     D -- 76128 Karlsruhe, Germany.} 
\end{center}

\vspace{\fill}

\begin{abstract}
\noindent
We suggest a method to determine $V_{ub} / V_{ts}$ through a 
comparison of  $B \to X_u \ell \bar{\nu}_\ell$ and 
$B \to X_s \gamma$. The relevant quantity is the spectrum of 
the light-cone component of the final state hadronic momentum, 
which in $B \to X_s \gamma$ is the photon energy while in 
$B \to X_u \ell \bar{\nu}_\ell$ this requires a measurement of 
both hadronic energy and hadronic invariant mass. 
The non-perturbative contributions at tree level to these distributions
are identical and may be cancelled by taking the ratio of the 
spectra. Radiative corrections to this comparison are discussed to 
order $\alpha_s$ and are combined with the non-perturbative 
contributions. 
\end{abstract}
\end{titlepage}

%
%
%
%
\section{Introduction}
$B$ meson decays with non-charmed final states will play an important role 
in the detailed investigations at future $B$--factories. Among these 
processes the ones involving the quark transitions 
$b \to u \ell \bar{\nu}_\ell$ and $b \to s \gamma$ are of interest
with respect to the determination of $V_{ub}$ and $V_{ts}$ as well 
as to discover or constrain effects of physics beyond the Standard 
Model (SM). 

From the theoretical side a lot of progress has been made by employing 
an expansion in inverse powers of the heavy quark mass $m_b$. Using 
operator product expansion (OPE) and the symmetries of Heavy Quark 
Effective Theory (HQET) \cite{HQET} the nonperturbative 
uncertainties can be reduced 
to a large extent in inclusive decays \cite{inclusive}. Concerning 
the transitions mentioned above the inclusive semileptonic or radiative 
processes such as $B \to X_u \ell \bar{\nu}_\ell$ and $B \to X_s \gamma$ 
are the ones which we shall address in the present paper. 

Looking at decay spectra of the leptons and photons in these processes, 
it has been pointed out that 
in certain regions of phase space (such as the endpoint region of the 
lepton energy spectrum in $B \to X_u \ell \bar{\nu}_\ell$ or the photon 
energy spectrum in $B \to X_s \gamma$ close to the endpoint of 
maximal energy) the correct description requires more than 
the naive $1/m_b$ expansion \cite{Neubert,Bigietal,MannelNeubert}. 
In these endpoint regions 
it is required to sum the leading twist 
contributions into a non-perturbative ``shape'' function, which describes the 
distribution of the light-cone component of the $b$ quark residual 
momentum inside the $B$ meson. 

Unfortunately, there is no way to avoid these endpoint regions due to 
experimental cuts. In $B \to X_u \ell \bar{\nu}_\ell$ cuts on the lepton 
energy and/or on the hadronic invariant mass of the final state are 
required to suppress the much larger charm contribution 
\cite{BaBarBook,Uraltsev,Falk}. Likewise, in 
$B \to X_s \gamma$ a cut on the photon energy is mandatory to reduce the 
background from ordinary bremsstrahlung in inclusive $B$ decays 
\cite{bsgammaexp}.
These cuts more or less reduce the accessible part of phase space to the 
endpoint regions. Hence it is unavoidable to get a theoretical handle 
on this light-cone distribution function.  
 
The distribution function is universal, since it depends only on the 
properties of the initial-state $B$ hadron. This fact may be used to 
establish a model independent relation between the two inclusive decays 
$B \to X_u \ell \bar{\nu}_\ell$ and $B \to X_s \gamma$ \cite{Neubert1}.
Parametrizations of this function have been proposed, which include 
the known features such as the few lowest moments 
\cite{BaBarBook,MannelNeubert}. It has also been 
shown that the popular ACCMM model \cite{ACCMM}
for a certain range of its parameters
is indeed consistent with this QCD based approach \cite{BigiUraltsev}.  

There have been various suggestions how to overcome the non-perturbative 
uncertainties induced by the distribution function. One way is to 
consider moments of appropriate distributions 
\cite{Neubert,Bigietal,MannelNeubert,Bauer,WiseLigeti}.
The first few moments 
are then sensitive only to the first few moments of the distribution 
function which are known. However, due to experimental cuts only parts 
of the distributions can be measured such that only moments involving  
cuts can be obtained. Depending on the cut, 
these quantities are sensitive to large 
moments of the distribution function and in this way the non-perturbative 
uncertainties reappear. 

In the present paper we propose a different approach. We suggest to directly 
compare the light-cone spectra of the final state hadrons in  
$B \to X_u \ell \bar{\nu}_\ell$ and $B \to X_s \gamma$. The individual 
rates depend on the shape function while the ratio of the two spectra 
is not very sensitive to non-perturbative effects 
even if cuts are included.
We discuss the kinematic variables which allow a direct 
comparison of the two processes and  consider the radiative 
corrections entering their relation. 
 
In the next section we give a 
derivation of the non-perturbative light cone distribution. Based on this 
we  calculate in section 3 the perturbative corrections and combine these 
with the non-perturbative light cone distribution function. Finally we 
study the comparison between $B \to X_u \ell \bar{\nu}_\ell$ and 
$B \to X_s \gamma$ quantitatively and conclude.

\section{Non-perturbative Contributions: 
         Light cone distribution function of the heavy quark}
Although the derivation of how the light-cone distribution function 
emerges in the heavy-to-light transitions at hand is well known 
\cite{Neubert,MannelNeubert,Bigietal}, we consider it useful to rederive it 
here, since our discussion of radiative corrections will be based on 
this derivation. 
We are going to consider heavy to light transitions such as $b \to u$ 
and $b \to s$ decays. The relevant effective Hamiltonian is obtained from 
the Standard Model by integrating out the top quark and the weak bosons.  
The dominant QCD corrections are taken into account as usual by 
running this effective interaction down to the scale of the bottom 
quark. The corresponding expressions are known to next-to-leading order 
accuracy \cite{Buras,AliGreub,ChetMM} and one obtains for the relevant pieces
\begin{eqnarray}
&& H_{eff}^{\rm sl} = \frac{G_F}{\sqrt{2}} 
(\bar{b} \gamma_\mu (1-\gamma_5) u)(\bar{\nu} \gamma_\mu (1-\gamma_5) \ell)
\\ 
&& H_{eff}^{rare} = \frac{G_F}{\sqrt{2}} \frac{e}{8 \pi^2} C_7 (\mu) 
m_b (\mu) (\bar{b} \sigma_{\mu \nu} (1-\gamma_5) s)|_\mu  F^{\mu \nu} 
\end{eqnarray}
mediating semileptonic $b \to u$ transitions and
radiative $b \to s$ decays respectively. $C_7 (\mu)$ is a Wilson 
coefficient, depending on the renormalization scale $\mu$, 
and $F^{\mu \nu}$ is the electromagnetic field strength. 

In the following we shall consider only these two contributions; at 
next-to-leading order (which is the accuracy we need here) there are
also contributions of other operators in the effective Hamiltonian. 
However, these contributions are small and can be neglected 
here \cite{WiseLigeti}, although they should be included when
analyzing experimental data.  

Considering the inclusive $B$ decays mediated by these two effective 
interactions, we shall look at the generic quantity 
\begin{equation}
R = \sum_X (2\pi)^4 \delta^4 (p_B - p_X - q) 
\langle B(p_B)| \bar{b} \Gamma q | X (p_X) \rangle  
\langle X (p_X)| \bar{q} \Gamma^\dagger b | B(p_B) \rangle 
\end{equation}
where $q$ is the momentum transferred to the non-hadronic system and 
$\Gamma$ is either $\gamma_\mu (1-\gamma_5)$ or $\sigma_{\mu \nu}$. 
 
We first redefine the phase of the heavy quark field 
\begin{equation}
b(x) = \exp(-im_b v \cdot x) b_v (x) 
\end{equation}
where $v$ is the velocity of the decaying $B$ meson 
\begin{equation} \label{velo}
v = \frac{p_B}{M_B} \, ;
\end{equation}
this corresponds to a splitting of the $b$ quark momentum according to 
$p_b = m_b v + k$, where $k$ is a small residual momentum. Going through 
the usual steps, we can rewrite $R$ as 
\begin{equation}
R = \int d^4 x \exp(-ix[m_b v - q])
\langle B(p_B)| \bar{b}_v (0) \Gamma q(0) 
\bar{q} (x) \Gamma^\dagger b_v(x) | B(p_B) \rangle 
\end{equation}
The momentum $P=m_b v -q$ is the momentum of the final state partons,
which is considered to be large compared to to any of the 
scales appearing in the matrix element. This allows us to set up
an operator product expansion (OPE).
If we assume 
\begin{equation}
(m_b v - q)^2 = {\cal O}(m^2) \mbox{ and } (m_b - v q) = {\cal O}(m)
\end{equation}
the OPE is a short distance expansion yielding the usual $1/m_b$ 
expansion of the rates. 

Close to the endpoint, where $P=m_b v - q$ is a practically 
light-like vector, which has still large components, i.e.
\begin{equation} \label{kinema}
(m_b v - q)^2 = {\cal O}(\Lambda_{QCD} m_b) 
\mbox{ and } (m_b - v q) = {\cal O}(m_b)
\end{equation}
one has to switch to a light cone expansion very similar to what 
is known in deep inelastic scattering. 

In order to study the latter case in some more detail, it is useful 
to define light-cone vectors as 
\begin{equation}
n_\pm = \frac{1}{\sqrt{(vP)^2 - P^2}} \left[ 
 v \left(\sqrt{(vP)^2 - P^2} \mp (vP)\right) \pm P \right]
\quad n_\pm^2 = 0 
\end{equation}
which satisfy the relations $n_+ n_- = 2$, $vn_\pm = 1$ and 
$2v = n_+ + n_-$. These vectors allow us to 
write 
\begin{equation}
P  = \frac{1}{2} ( P_+ n_- + P_- n_+ ) \qquad P_\pm  = P n_\pm 
\end{equation} 
where 
\begin{eqnarray}
&& P_+ = (vP) - \sqrt{(vP)^2 - P^2} 
       \to - \frac{P^2}{2 (vP)} \mbox { for } P^2 \ll (vP)^2 \\
&& P_- = (vP) + \sqrt{(vP)^2 - P^2} 
       \to 2 (vP) \mbox { for } P^2 \ll (vP)^2 
\end{eqnarray}
The kinematic region in which the light cone distribution function 
becomes relevant is the one where $P_+$ is much smaller than $P_-$:
Here the main contributions
to the integral come from the light cone $x^2 \approx 0$, since 
in order to have a contribution to the integral we need to have 
\begin{equation}
x\cdot P = \frac{1}{2} (x_- P_+ + x_+ P_-) = x_+ (vP) + \mbox{ const}  
< \infty 
\end{equation}
in the limit in which $ P_- \to \infty$. Hence $x_+$ has to be small, 
restricting the integration to the light cone. 

We consider first the tree level contribution, 
which corresponds to a contraction of the light quark line; 
we get \cite{MannelNeubert}
\begin{eqnarray}
R &=& \int d^4 x \int \frac{d^4 Q}{(2\pi)^4} \Theta(Q_0) (2\pi)
\delta(Q^2) \exp(-ix[m_b v - q - Q]) \\ \nonumber 
&& \langle B(v)| \bar{b}_v (0) \Gamma \fmslash{Q} 
                \Gamma^\dagger b_v(x) | B(v) \rangle 
\end{eqnarray}
where now $|B(v)\rangle$ is the static $B$ meson state. 
Performing a (gauge covariant) Taylor expansion of the 
remaining $x$ dependence of the matrix element, we get
\begin{eqnarray} \label{taylor}
R &=& \int d^4 x \int \frac{d^4 Q}{(2\pi)^4} \Theta(Q_0) (2\pi)
\delta(Q^2) \exp(-ix[m_b v - q - Q]) \\ \nonumber 
&& \sum_n \frac{1}{n!} \langle B(v)| \bar{b}_v (0) \Gamma \fmslash{Q} 
                \Gamma^\dagger (-ix \cdot iD)^nb_v(0) | B(v) \rangle 
\\ \nonumber
&=& \int d^4 x \int \frac{d^4 Q}{(2\pi)^4} \Theta(Q_0) (2\pi)
\delta(Q^2) \\ \nonumber 
&& \langle B(v)| \bar{b}_v (0) \Gamma \fmslash{Q} 
                \Gamma^\dagger {\cal P} \exp(-ix[m_b v - q - Q + iD]) 
   b_v(0) | B(v) \rangle 
\end{eqnarray}
where the symbol ${\cal P}$ means the usual path--ordering of the exponential.
Using spin symmetry and the usual representation matrices of the
$0^-$ B meson states,  
the matrix element which appears in (\ref{taylor}) becomes 
\begin{eqnarray}
&& \langle B(v)| \bar{b}_v (0) \Gamma \fmslash{Q} 
\Gamma^\dagger 
(iD_{\mu_1}) (iD_{\mu_2}) \cdots (iD_{\mu_n}) b_v(0) | B(v) \rangle 
\\ \nonumber
&& = \frac{M_B}{2} \mbox{Tr}\{\gamma_5 (\fmslash{v}+1) \Gamma \fmslash{Q} 
 \Gamma^\dagger (\fmslash{v}+1)\gamma_5 \}
 [a_1^{(n)} v_{\mu_1}  v_{\mu_2} \cdots v_{\mu_n}  
 + a_2^{(n)} g_{\mu_1 \mu_2} v_{\mu_3} \cdots v_{\mu_n} + \cdots ]
\end{eqnarray}
where the elipses denote terms with one or more $g_{\mu \nu}$'s and 
also antisymmetric terms and $a_i^{(n)}$ are non-pertrubative parameters.
Contracting with 
$x^{\mu_1}  x^{\mu_2} \cdots x^{\mu_n}$ all antisymmetric contributions 
vanish; furthermore, since the relevant kinematics restricts the $x_\mu$ 
to be on the light cone, also all the $g_{\mu \nu}$ terms are suppressed 
relative to the first term, which has only $v_\mu$'s. Hence
\begin{equation} \label{term}
\langle B(v)| \bar{b}_v (0) \Gamma \fmslash{Q} 
                \Gamma^\dagger (-ix \cdot iD)^n b_v(0) | B(v) \rangle 
= M_B \mbox{Tr}\{(\fmslash{v}+1) \Gamma \fmslash{Q} \Gamma^\dagger \}
  a_1^{(n)} (v\cdot x)^n
\end{equation}
In this way we have made explicit that the kinematics we are studying 
here forces us to resum the series in $1/m_b$. Defining the twist 
$t$ of an operator ${\cal O}$ in the usual way 
$t = dim[{\cal O}] - \ell$, where $\ell$ is the spin of the operator, we 
find that the resummation corresponds to the contributions of leading 
twist, $t=3$. 
Since $x_\mu$ is light like, it projects out only the light cone 
component $D_+$ of the covariant derivative in (\ref{term}). 
Hence we may write $a_1^{(n)}$ as 
\begin{equation}
2 M_B a_1^{(n)} = \langle B(v)| \bar{b}_v (iD_+)^n b_v  | B(v) \rangle 
\end{equation}
and one obtains as a final result 
\begin{eqnarray}
R &=& \int d^4 x \int \frac{d^4 Q}{(2\pi)^4} \Theta(Q_0) (2\pi)
\delta(Q^2) \frac{1}{2} 
\mbox{Tr}\{(\fmslash{v}+1) \Gamma \fmslash{Q} \Gamma^\dagger \}
\\ \nonumber 
&& \langle B(v)| \bar{b}_v \exp(-ix[(m_b+iD_+) v - q - Q]) 
   b_v | B(v) \rangle 
\end{eqnarray}
Introducing the shape function (or light cone distribution function)
as \cite{Neubert,MannelNeubert,Bigietal} 
\begin{equation}
2 M_B f(k_+) = \langle B(v)| \bar{b}_v \delta(k_+ - iD_+) b_v | B(v) \rangle  
\end{equation}
we can write the result as 
\begin{eqnarray} \label{final}
R &=& \int dk_+ f(k_+) \int \frac{d^4 Q}{(2\pi)^4} \Theta(Q_0) (2\pi)
\delta(Q^2) \\ \nonumber && M_B
\mbox{Tr}\{(\fmslash{v}+1) \Gamma \fmslash{Q} \Gamma^\dagger \}
(2 \pi)^4 \delta^4 ([m_b + k_+]v -Q - q)
\end{eqnarray}
This result shows that the leading twist contribution is obtained 
by convoluting the partonic result with the shape function, where 
in the partonic result the $b$ quark mass $m_b$ is replaced by 
the mass 
\begin{equation} \label{mstar}
m_b^* = m_b + k_+
\end{equation}  

Let us take a closer look at the kinematics. The final-state quark 
is massless, 
$$
0 = ([m_b+k_+] v - q)^2 
$$
in which case we have for  
the light cone component of the heavy quark residual momentum\footnote{%
    There are actually two solutions, but only one vanishes 
    at the endpoint}
\begin{equation} \label{kplus}
k_+ = -m_b + (vq) + \sqrt{(vq)^2 - q^2} 
\end{equation}
Note that this variable -- up to a minus sign -- is just $P_+$ and 
thus close to the endpoint
\begin{equation}
k_+ \approx - \frac{P^2}{2(vP)} 
\end{equation}

If we now look at the hadronic kinematics 
\begin{equation}
M_B v - q = p_X \, ,
\end{equation} 
where $p_X$ is the four momentum of the final state hadrons, 
and use the relation 
between the heavy quark mass and the meson mass 
\begin{equation} \label{mass} 
M_B = m_b + \bar{\Lambda} + {\cal O}(1/m_b) 
\end{equation}
we find 
\begin{equation} 
p_X = P + \bar{\Lambda} v \quad \mbox{ or } \quad 
      p_{X\pm} = P_\pm + \bar{\Lambda}
\end{equation}

Thus the light cone component 
of the hadronic momentum of the final state 
\begin{equation}
L_+ = - p_{X+} = - M_B + (vq) + \sqrt{(vq)^2 - q^2} 
\end{equation}
ranging between $-M_B \le L_+ \le 0$ is directly related to 
the light-cone component of the residual momentum of the heavy 
quark
\begin{equation}
L_+ = k_+ - \bar{\Lambda}
\end{equation}
where the range of $k_+$ is given by $-m_b \le k_+ \le \bar{\Lambda}$.
However, large negative values of $k_+$ close to $-m_b$ are beyond 
the validity of the heavy mass limit. 

The observable $L_+$ directly measures the light cone component 
of the residual momentum of the heavy quark, at least for small values  
of $L_+$. In the case of $b \to s \gamma$ we have 
$q^2 = 0$ and $L_+$ is directly related to the energy $E_\gamma$ of the photon 
in the rest frame of the decaying $B$ 
$$
L_+ = - M_B + 2 E_\gamma .
$$ 
For 
$b \to u \ell \bar{\nu}_\ell$ a reconstruction of $L_+$ requires both
a measurement of the hadronic energy and the hadronic invariant mass.

Reexpressing the 
convolution (\ref{final}) in terms of $L_+$ and using that the
light-cone distribution function is non-vanishing only for 
$-m_b \to - \infty < k_+ < \bar{\Lambda}$
we may rewrite (\ref{final}) as 
\begin{eqnarray} \label{final1}
R &=& \int\limits_{-M_B}^0 dL_+ f(L_+ +\bar{\Lambda}) 
\int \frac{d^4 Q}{(2\pi)^4} \Theta(Q_0) (2\pi)
\delta(Q^2) \\ \nonumber && M_B
\mbox{Tr}\{(\fmslash{v}+1) \Gamma \fmslash{Q} \Gamma^\dagger \}
(2 \pi)^4 \delta^4 ([M_B + L_+]v -Q - q)
\end{eqnarray}
where we have made use of (\ref{mass}). Equivalently, we may write 
the spectrum in the variable $L_+$ as 
\begin{eqnarray} \label{final2}
\frac{dR}{dL_+} &=& f(L_++\bar{\Lambda}) 
\int \frac{d^4 Q}{(2\pi)^4} \Theta(Q_0) (2\pi)
\delta(Q^2) \\ \nonumber && M_B
\mbox{Tr}\{(\fmslash{v}+1) \Gamma \fmslash{Q} \Gamma^\dagger \}
(2 \pi)^4 \delta^4 ([M_B + L_+]v -Q - q)
\end{eqnarray}

The two relations (\ref{final1}) and (\ref{final2}) exhibit an 
interesting feature of the summation of the leading twist terms, namely 
that the dependence on the heavy quark mass has completely disappeared,
only the shape function $f$ still depends on $\bar\Lambda$. The full result 
should be independent of this unphysical quantity and hence the  
explicit dependence on $\bar{\Lambda}$ of the shape 
function has to cancel against the one appearing in the argument of the 
shape function. In other words, if we write the shape function as 
$f = f(k_+,\bar{\Lambda})$, this function of two variables has to satisfy
\begin{equation} \label{Labindependence}
\left(\frac{\partial}{\partial k_+} + 
      \frac{\partial}{\partial \bar{\Lambda}} \right) f(k_+,\bar{\Lambda}) 
= 0 \quad \mbox{ or } \quad f(k_+,\bar{\Lambda}) = g(k_+ - \bar{\Lambda}) 
\end{equation} 
In this way the total result becomes independent of any reference to the 
quark mass $m_b$ or equivalently on $\bar{\Lambda}$. In particular,
(\ref{Labindependence}) ensures the cancellation 
of ambiguities related to the renormalon in the heavy quark mass
\cite{BigiUraltsev}. 

Putting everything together, the 
light-cone spectrum (\ref{final2}) for the decay 
$B \to X_u \ell \bar{\nu}_\ell$ becomes
\begin{equation} \label{final3}
\frac{d \Gamma}{dL_+} (B \to X_u \ell \bar{\nu}_\ell) =  
\frac{G_F^2 (M_B + L_+)^5 |V_{ub}|^2}{192 \pi^3} g(L_+) 
\end{equation}
which means that the spectrum is proportional to the total rate 
in which the $m_b$ has been replaced by $M_B + K_+$. 

Similarly, for $B \to X_s \gamma$ one finds 
\begin{equation} \label{final4}
\frac{d \Gamma}{dL_+} (B \to X_s \gamma)= 
\frac{G_F^2 (M_B + L_+)^5 |V_{tb} V_{ts}^*|^2 \alpha_{em} }{32 \pi^4} 
[C_7 (M_B)]^2 g(L_+) 
\end{equation}

Note that (\ref{final3}) and (\ref{final4}) do not depend on 
any unphysical parameter, 
since the $\bar\Lambda$ dependence has disappeared. In particular, 
all ambiguities induced by renormalons should cancel in these equations. 
Based on this it has been suggested to give a definition for a 
heavy quark mass $\hat{m}_b$ through (\ref{final3}) and (\ref{final4}) 
using e.g. the mean energy $\langle E_\gamma \rangle$ of the  photon in 
$B \to X_s \gamma$ \cite{Bauer}. To this end we consider 
the moments of the function $g$, which can be obtained from the 
moments of the shape function $f$. The usual HQET relations are
\begin{eqnarray} \label{moments}
&& \int_{-\infty}^{\bar\Lambda} dk_+ f(k_+) = 1 \label{zero}\\
&& \int_{-\infty}^{\bar\Lambda} dk_+ k_+ f(k_+) = \delta m_b \label{one}\\
&& \int_{-\infty}^{\bar\Lambda} dk_+ k_+^2 f(k_+) \label{two}
= -\frac{1}{3} \lambda_1 
\end{eqnarray}
which define the normalization, the residual mass term \cite{FalkNeubert}
and the kinetic energy parameter. 

Reexpressing relation (\ref{one}) through the function $g$ we have
\begin{equation}
\int_{-\infty}^0 dL_+ L_+  g(L_+) = \delta m_b - \bar\Lambda 
\equiv \hat\Lambda 
\end{equation}
which gives a definition of the heavy quark mass 
$\hat{m_b} = M_B - \hat\Lambda$ free of renomalon ambiguities which 
cancel between the $\delta m_b$ and $\bar\Lambda$ \cite{Bauer}.

For the second moment one obtains from (\ref{two}) 
\begin{equation} \label{two1}
\int_{-\infty}^0 dL_+ L_+^2 g(L_+) 
= -\frac{1}{3} \lambda_1 - \bar\Lambda^2 - 2 \bar\Lambda \hat\Lambda
\equiv -\frac{1}{3} \hat\lambda_1
\end{equation}
where $\hat\lambda_1$ again has to be free of renormalon ambiguities.

We shall later need a model shape function to discuss our results. 
In order to keep things simple, we use the one-parameter 
model of \cite{MannelNeubert} 
\begin{eqnarray} \label{sf}
g(L_+) &=& {32 \over \pi^2 \sigma^3} L_+^2 
     \exp\left[- {4 \over \pi \sigma^2 } L_+^2 \right]
    \Theta(-L_+) 
\end{eqnarray}
which is easy to deal with. Here $\sigma$ is the width 
parameter of the shape function and will be varied in our numerical 
studies between 400 MeV and 800 MeV.

\section{Including Perturbative Contributions} 
The next step is to include radiative corrections to the relations 
(\ref{final3}) and (\ref{final4}). The starting point of the 
considerations is (\ref{final}) or eqivalently (\ref{final1}). 
However, now the partonic result $d\Gamma_{part}$ including 
the radiative corrections is convoluted with the shape 
function
\begin{equation}
d\Gamma_{had} = \int\limits_{-m_b}^{\bar{\Lambda}} 
           d\Gamma_{part} (m_b^* =  m_b + k_+) f(k_+) dk_+
\end{equation}
We shall compute the spectrum in the light cone variable of the 
hadronic momentum $L_+$, and the partonic counterpart of this 
variable is 
\begin{equation}
l_+ = - m_b + (vq) + \sqrt{(vq)^2 - q^2} \, .
\end{equation}
in which we compute a differential spectrum as some function $G$ of 
$l_+$ and $m_b$ 
\begin{equation}
\frac{d\Gamma_{part}}{d l_+} = G(m_b,l_+) \, .
\end{equation}
To apply the convolution formulae (\ref{final}) and (\ref{final1}) 
we replace in the first step the heavy quark mass by $m_b^*$
and convolute with the shape function. The variable $l_+$  
is also replaced by
\begin{equation}
l_+^* = - m_b^* + (vq) + \sqrt{(vq)^2 - q^2} = l_+ - k_+
\end{equation}    
Note that $l_+^*$ is again a light cone variable for a process in which 
the heavy quark mass is replaced by $m_b^*$, and hence it ranges between
$-m_b^* \le l_+^* \le 0$, which means that the $k_+$ integration becomes 
restricted to 
\begin{equation}
-l_+ \le k_+ \le \bar{\Lambda}
\end{equation} 

The second step towards hadronic variables is to replace $l_+$ by $L_+$, 
the hadronic light cone momentum. At the same time we perform a shift 
in the integration variable $k_+ \to K_+ - \bar{\Lambda}$ and obtain 
\begin{equation} \label{pert}
\frac{d\Gamma_{had}}{dL_+} =  \int\limits_{L_+}^0 dK_+ \,  
             G(M_B+K_+,L_+ -K_+) g(K_+)  
\end{equation}

Relation (\ref{pert}) holds for any $b \to q$ transition where $q$ is 
a light quark. We shall exploit (\ref{pert}) for a comparison between 
the inclusive processes $B \to X_s \gamma$ and 
$B \to X_u \ell \bar{\nu}_\ell$. The partonic $l_+$ spectra of the two 
processes can be calculated; both take the generic form
\begin{eqnarray} \label{partonic}
G(m_b,l_+) &=& \gamma_0 m_b^5 
\left[ (1+ b_0 \frac{\alpha_s}{3\pi}) \delta(l_+) +
\vphantom{\int\limits_t^t} \right. \\ \nonumber 
&& \left. \frac{\alpha_s}{3\pi} \left( 
b_1 \left(\frac{\ln(-l_+ / m_b)}{-l_+} \right)_+ + 
b_2 \left(\frac{1}{-l_+} \right)_+ 
+ \frac{1}{m_b} \Phi \left(\frac{-l_+}{m_b} \right) \right) \right]
\end{eqnarray}
where we have defined the ``+''-distributions in the usual way
\begin{equation}
\int_0^1 dx D_+ (x) = 0 \, , 
\end{equation}
and $b_0$, $b_1$, $b_2$ and $\Phi$ are known quantities. 
For $b\to u \ell \bar{\nu}_\ell$ we find 
\begin{eqnarray} \label{b2uln}
\gamma_0 &=& {G_F^2 \left|V_{ub}\right|^2 \over 192 \pi^3} \\
b_0      &=& -{13\over 36}-2\pi^2 \nonumber \\
b_1      &=& -4 \nonumber\\
b_2      &=& -26/3 \nonumber \\
\Phi(x)  &=& \frac{158}{9} + \frac{407\,x}{18} - \frac{367\,{x^2}}{6} +
\frac{118\,{x^3}}{3}
                 - \frac{100\,{x^4}}{9} + \frac{11\,{x^5}}{6} -
\frac{7\,{x^6}}{18} \nonumber \\
&&               - \frac{4\,\ln (x)}{3} + \frac{46\,x\,\ln (x)}{3} +
6\,{x^2}\,\ln (x) 
                - \frac{16\,{x^3}\,\ln (x)}{3} \nonumber\\ 
&&              - 12\,{x^2}\,{{\ln (x)}^2} + 8\,{x^3}\,{{\ln (x)}^2}
\nonumber
\end{eqnarray}

While for $b\to s\gamma$
\begin{eqnarray} \label{b2sg}
\gamma_0 &=& {G_F^2 \left|V_{ts}V_{tb}^*\right|^2 \left| C_7\right|^2
             \alpha_{\rm em} \over 32 \pi^4} \\
b_0      &=& -5-\frac{4}{3} \pi^2 \nonumber \\   
b_1      &=& -4 \nonumber \\
b_2      &=& -7 \nonumber \\
\Phi(x)  &=& 6 + 3\,x - 2\,{x^2} - 4\,\ln (x) + 2\,x\,\ln (x) \nonumber
\end{eqnarray}
where $C_7 = -0.306$ is obtained from the next-to-leading order calculation 
of \cite{ChetMM}. Our result for $B \to X_s \gamma$ coincides with the ones 
obtained in the literature \cite{AliGreub,ChetMM,KaganNeubert}. 
To determine $b_0$ for $B \to X_u \ell \bar{\nu}_\ell$ and $B \to X_s \gamma$
the respective total rates at ${\cal O}(\alpha_s)$ \cite{Timo,Pott} have
been used.

The partonic result up to order $\alpha_s$ contains terms of all 
orders in the $1/m_b$ expansion and we shall first consider the 
leading twist contribution. Taking the 
limit $m_b \to \infty$ of the partonic calculation leaves us only 
with the ``+''-distributions and the $\delta$ function. The 
convolution becomes 
\begin{eqnarray} 
\frac{d\Gamma_{had}}{dL_+} &=&  \int\limits_{L_+}^0 dK_+ \, g(K_+)    
\gamma_0 (M_B+K_+)^5 
\left[ (1+ b_0 \frac{\alpha_s}{3\pi}) \delta(L_+-K_+) 
\vphantom{\int\limits_t^t} \right. \\ \nonumber 
&& + \frac{\alpha_s}{3\pi} \left. \left( 
b_1 \left(
\frac{\ln\left(\frac{-L_+ + K_+}{M_B + K_+}\right)}{-L_+ + K_+} \right)_+ + 
b_2 \left(\frac{1}{-L_+ + K_+} \right)_+ \right) \right]
\end{eqnarray}
The shape function $g(K_+)$ is restricted to values $K_+$ small 
compared to $M_B$, and one may expand the dependence on $M_B + K_+$ 
for small $K_+$. This induces terms of higher orders in $1/m_b$, which 
again may be dropped. Hence the leading twist terms become
\begin{eqnarray}
\frac{d\Gamma_{had}}{dL_+} &=&  \gamma_0 M_B^5  
\left[ (1+ b_0 \frac{\alpha_s}{3\pi}) g(L_+) +
\vphantom{\int\limits_t^t} \right. \\ \nonumber 
&&  \left. \frac{\alpha_s}{3\pi} \int\limits_{L_+}^0 dK_+ g(K_+) \left( 
b_1 \left(
\frac{\ln\left(\frac{-L_+ + K_+}{M_B}\right)}{-L_+ + K_+} \right)_+ + 
b_2 \left(\frac{1}{-L_+ + K_+} \right)_+ \right) \right] 
\end{eqnarray}

It is well known that the coefficient $b_1$ is universal as well as 
the shape function.
This suggests to define a scale dependent shape function, which to 
order $\alpha_s$ becomes
\begin{equation} \label{scale}
{\cal F}(K_+,\mu) = g(K_+) + \frac{\alpha_s b_1}{3 \pi}  
\int\limits_{L_+}^0 dK_+ g(K_+) \left( 
\frac{\ln\left(\frac{-L_+ + K_+}{\mu}\right)}{-L_+ + K_+} \right)_+ 
\end{equation}
The scale dependence of the distribution function has been discussed 
in \cite{BMK}, where the  evolution equation for the distribution function 
has been set up.

The terms proportional to $(1/l_+)_+$ are not universal, 
which is natural, since the relavant scale in two different processes 
may be different. A change of scale changes the contribution
of the $(\alpha_s / \pi) (1/l_+)_+$ pieces according to 
\begin{equation}
{\cal F}(K_+,\mu) = {\cal F}(K_+,\mu^\prime) + 
\frac{\alpha_s b_1}{3 \pi}  
\int\limits_{L_+}^0 dK_+ g(K_+)  
\ln\left(\frac{\mu^\prime}{\mu}\right)
\left(\frac{1}{-L_+ + K_+} \right)_+ 
\end{equation}
and hence we find for the leading twist contribution
\begin{equation} \label{part1}
\frac{d\Gamma_{had}}{dL_+} =  \gamma_0 M_B^5   
(1+ b_0 \frac{\alpha_s}{3\pi}) {\cal F}(L_+,\mu)
\end{equation}
where the scale $\mu$ is given by 
\begin{equation} \label{scalemu}
\ln\left(\frac{M_B}{\mu}\right) = \frac{b_2}{b_1} 
\end{equation} 
Inserting the results for the two processes under consideration 
we find 
\begin{equation} \label{scales}
\mu_{b \to s \gamma} = 0.1738 M_b \qquad 
\mu_{b \to u \ell \bar{\nu}_\ell} = 0.1146 M_b 
\end{equation}
which are in fact scales small compared to the mass of the $b$ quark. 
Furthermore, the ratio of the two scale is of order one, 
\begin{equation}
\frac{\mu_{b \to s \gamma}}{\mu_{b \to u \ell \bar{\nu}_\ell}} \approx 
1.52
\end{equation}
The physical meaning of the scale $\mu$ can be understood in a picture 
as proposed in \cite{KorchemskySterman}. Here the amplitude is decomposed 
into a hard part, a ``jetlike'' part incorporating the collinear 
singularities and a soft part. The typical scales 
corresponding to these pieces 
are $m_b^2$ for the hard, $\Lambda_{QCD} m_b$ for the ``jetlike'' and 
$\Lambda_{QCD}^2$ for the soft part. Comparing the present approach 
to \cite{KorchemskySterman} the light-cone distribution function 
corresponds to the convolution of the soft and the ``jetlike'' piece, 
both of which are universal but scale dependent. 

Thus the scale appearing in the shape function should be of the order 
$\mu^2 \sim m_b \Lambda_{QCD}$ since it incorporates all lower scales. 
In particular, the shape function combines both the collinear and the 
soft contributions which according to \cite{KorchemskySterman} could be 
factorized. Since we are working to 
next--to--leading order, we can fix the scale according to (\ref{scalemu}), 
and we expect to obtain a scale $\mu^2$ of order 
$\Lambda_{QCD} m_b \sim 0.9$ GeV${}^2$,  which is confirmed by (\ref{scales}).  

The rest of the partonic result, i.e. the function $\Phi$, 
is a contribution of subleading terms, suppressed by at least one 
power of the heavy mass. We shall use these terms to estimate, how 
far we can trust the leading twist terms, in particular in the 
comparison between 
$B \to X_u \ell \bar{\nu}_\ell$ and $B \to X_s \gamma$. 
To this end we add the two contributions and use the expression
\begin{equation} \label{partfull}
\frac{d\Gamma_{had}}{dL_+} =  \gamma_0 M_B^5 \left[   
(1+ b_0 \frac{\alpha_s}{3\pi}) {\cal F}(L_+,\mu) 
 + \frac{\alpha_s}{3\pi M_B} \Phi \left(-\frac{L_+}{M_B} \right) \right] 
\end{equation}
We note that the functions $\Phi$ for both processes contain terms 
diverging logarithmically as $L_+ \to 0$, but which are suppressed 
by $1/m_b$. These divergencies will be cured once the analogon of the 
light-cone distribution function at subleading order is taken into 
account. However, for the quantitative analyis of the process these 
terms are not important as long as we do not get too close to the 
endpoint.      

\section{Comparison of $B \to X_u \ell \bar{\nu}_\ell$ 
           to $B \to X_s \gamma$}
\begin{figure}
\hspace*{-15mm}
  \leavevmode
  \epsfxsize=8cm
  \epsffile{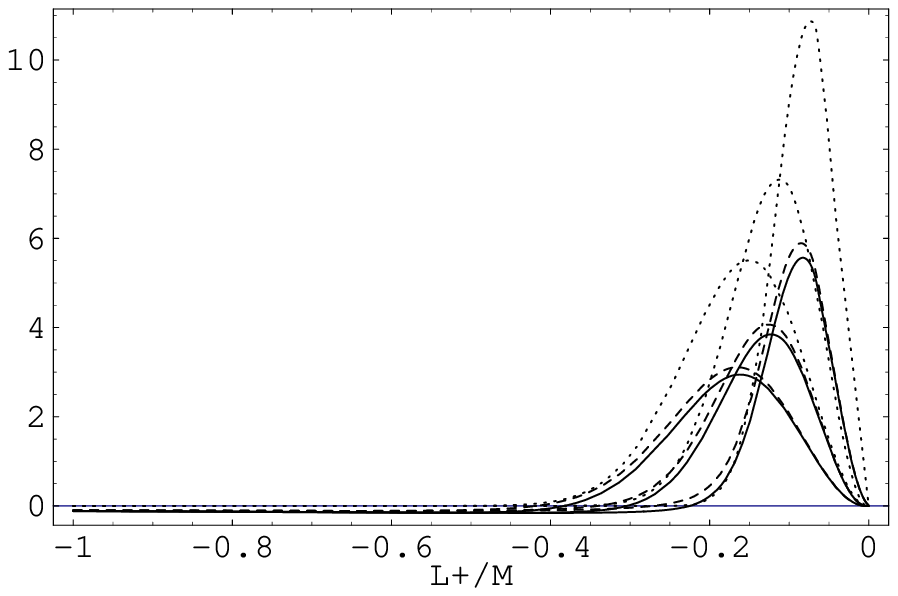}
  \epsfxsize=8cm
  \epsffile{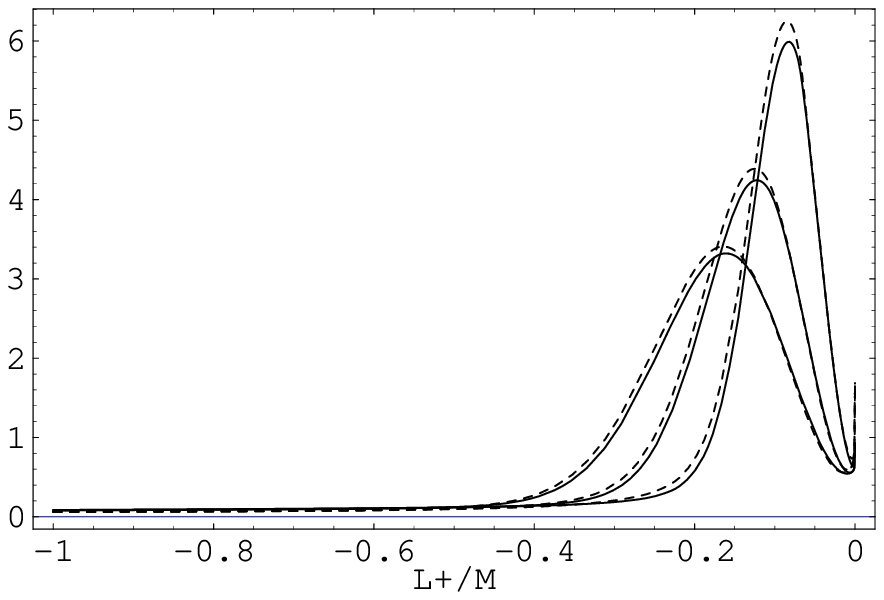}
\caption{Light-cone distribution spectra for 
         $B \to X_u \ell \bar{\nu}_\ell$ (solid) and 
         $B \to X_s \gamma $ (dashed) in units of $\gamma_0 M_B^5$. 
         The dotted line is the leading contribution, which is the 
         same for both processes. The three sets of curves correspond
         to values of $\sigma$ of 400, 600 and 800 MeV.}
\label{cfig1} 
\end{figure}

We shall first study the $L_+$ spectra of the two processes 
using (\ref{part1}) and (\ref{partfull}). These results depend on the 
model for the shape function one is using and we shall employ the simple 
one-parameter model (\ref{sf}). 
In fig.\ref{cfig1} we show the $L_+$ spectra of  
$B \to X_u \ell \bar{\nu}_\ell$ and $B \to X_s \gamma$. We divide the  
rates by the prefactors $\gamma_0 M_B^2$, such that at leading order
only the shape function remains. The left plot shows the shape
function (\ref{sf}) together with the 
leading twist contribution (\ref{part1}) for the two processes 
for three different values of the width parameter $\sigma$. 

The leading twist contribution can only be trusted close to the endpoint
$L_+ =0$; it even becomes negative for values 
$L_+ \le - \sigma$, indicating the 
breakdown of the leading twist approximation. Adding the subleading 
terms as in (\ref{partfull}) cures this problem and the results are 
shown in the right hand figure of fig.\ref{cfig1}. The sharp rise of the 
full results is due to the logarithms $\ln L_+$ of order $1/m_b$, which 
become relevant only very close to $L_+ = 0$. These contributions are 
integrable and will not be relevant once the rates are binned with 
reasonably large bins.   

Next we shall compare $B \to X_u \ell \bar{\nu}_\ell$ 
to $B \to X_s \gamma$ by considering the ratio of the two $L_+$ 
distributions. From the experimental point of view many systematic
uncertainties cancel, while from the theoretical side one 
expects to reduce nonperturbative uncertainties, which indeed 
at tree level cancel completely. 

The perturbative result for the spectrum is a distribution and 
the ratio of the perturbative expressions is meaningless, even if 
we avoid the region around $L_+ = 0$. However, once we have combined 
both perturbative and non-perturbative contributions we can take the 
ratio without encountering the problem of distributions. 

The price we have to pay is that the resulting ratio is not 
independent of the shape function, an effect which has 
been observed already in \cite{Neubert1}. Still the ratio of the 
two processes is a useful quantity, since one may expect that 
much of the non-perturbative uncertainties still cancel. 
 
\begin{figure}
\hspace*{-15mm}
  \leavevmode
  \epsfxsize=8cm
  \epsffile{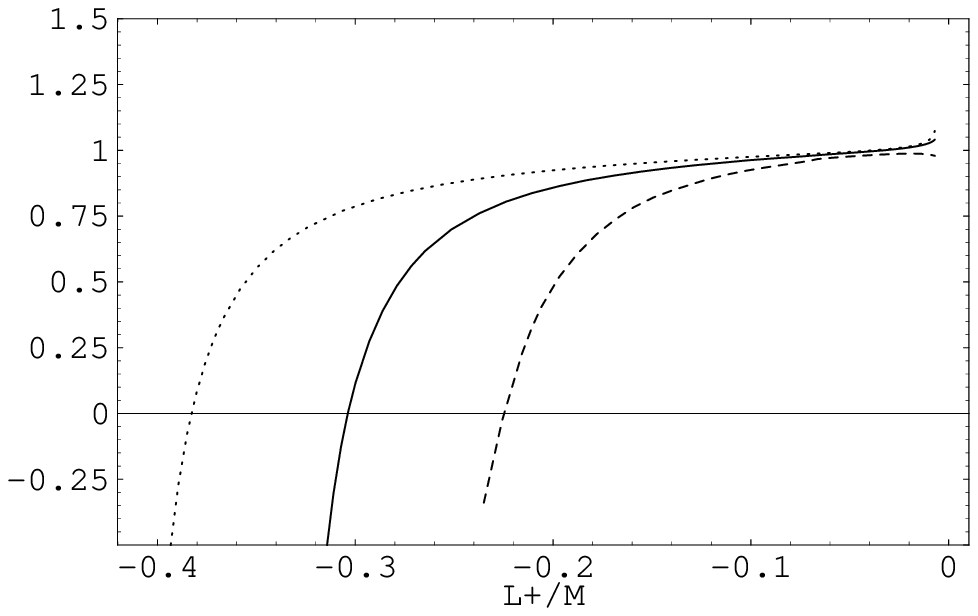} 
  \epsfxsize=8cm
  \epsffile{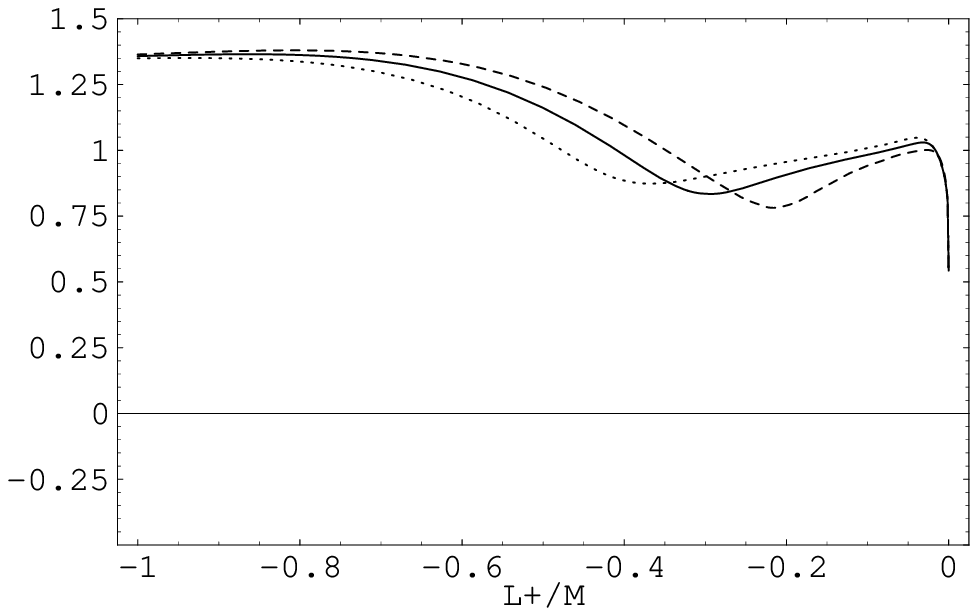}
\caption{Ratio of the light cone distribution spectra for the two 
         processes. The leading twist contribution (\protect{\ref{comp}})
         is given  in the left plot while the right plot shows 
         the full result.
         The solid, dashed and dotted lines correspond to values
         of $\sigma$ of 600, 400 and 800 MeV, respectively.} 
\label{cfig2}
\end{figure}

The two spectra are due to the ``smearing'' with the nonperturbative 
shape function smooth curves and we can perform a comparison of the 
two processes by taking the ratio. At leading twist this ratio becomes 
\begin{equation} \label{comp}
\frac{(d \Gamma_{B \to X_u \ell \bar{\nu}_\ell} / dL_+)}
     {(d \Gamma_{B \to X_s \gamma} / dL_+)}
= \frac{|V_{ub}|^2}{|V_{ts} V_{tb}^*|^2}\frac{\pi}{6 \alpha_{em} |C_7|^2}
  \frac{{\cal F}(L_+,\mu_{b \to s \ell \bar{\nu}_\ell})}
       {{\cal F}(L_+,\mu_{b \to s \gamma})}       
\left(1 + \frac{\alpha_s (m_b)}{3 \pi} \left[\frac{167}{36} - 
     \frac{2}{3} \pi^2 \right] \right)  
\end{equation}
where the scales of the distribution functions have been given 
in (\ref{scales}). It is interesting to note that although the radiative 
corrections at leading twist (the constants $b_0$) are big for the 
individual processes, they turn out to be small in the ratio (\ref{comp}).

For values of $L_+$ within the nonperturbative region 
$L_+ \ge -\sigma$ 
the leading twist contribution is dominant and the comparison may be 
performed using the leading twist terms only. This is the region with
the largest fraction of the total rate and hence the subleading terms 
will not play a role, also due to experimental cuts. The ratio of the 
leading twist terms is shown in the left hand plot of fig.\ref{cfig2}
for three values of the parameter $\bar\Lambda$. 

The region of validity of the leading twist contribution may be studied 
by looking at the ratio of the full results (\ref{partfull}) shown in 
fig(\ref{cfig2}). The subleading terms lead to a minimum at the point 
where they are becomming the dominant contribution to the rates. This 
happens at values of $L_+$ which correspond to the typical width $\sigma$ 
of the non-perturbative shape function.

For values of $L_+$ between $-\sigma$ and $0$ one may then use our 
results to determine CKM matrix elements, since the ratio of the 
two rates depends on 
$$
\frac{|V_{ub}|^2}{|V_{ts} V_{tb}^*|^2}\frac{\pi}{6 \alpha_{em} |C_7|^2} 
\approx 5
$$
i.e. the results shown in fig.\ref{cfig2} have to be multiplied by 
this factor. Hence one may determine $ |V_{ub}|/|V_{ts}|$ in this 
way; however, the radiative corrections induce a small hadronic 
uncertainty, which is hard to quantify without knowledge of the 
non-perturbative distribution function. Still one may expect only a 
small uncertainty, since at tree level there is no hadronic uncertainty 
at all.   

\section{Conclusions}
Due to experimental constraints severe cuts on the phase space in 
both $b \to s \gamma$ and $b \to u \ell \bar{\nu}_\ell$ inclusive 
transitions have to be imposed making these decays sensitive to 
non-perturbative effects. Within the framework of the heavy mass 
expansion of QCD these effects are encoded in a light-cone distribution 
function which is universal for all heavy to light processes. 

The universality of this function may be exploited to eliminate 
the non-perturbative uncertainties in $b \to s \gamma$ using the 
input of $b \to u \ell \bar{\nu}_\ell$ or vice versa. To this end, 
one may use the comparison of the two processes to determine the 
ratio $V_{ub} / V_{ts}$ or --- for fixed CKM parameters --- to test 
the Wilson coefficient $C_7$. 

The comparison between the two processes is best performed in terms 
of the spectra of the light-cone component of the final state hadrons,
which for the case of $B \to X_s \gamma$ is equivalent to 
the photon energy spectrum, while for $B \to X_u \ell \bar{\nu}_\ell$  
this requires a measurement of both the hadronic invariant mass and the 
hadronic energy. 

For small values, the light cone component of the final state hadrons is 
the same as the light cone component of the heavy quark residual momentum.  
The corresponding spectra are at tree level directly proportional to 
the light cone distribution function and hence one may access this 
function by a measurement of the the light cone distribution 
of the momentum of the final state hadrons. 

However, radiative corrections change this picture. The main effect is 
that the shape function does not cancel any more in the ratio of the 
two spectra. The 
partonic radiative corrections to the light-cone distribution of 
the final state partons exhibit the well known singularities of 
the type $\ln l_+ / l_+$ and $1/l_+$ which can be absorbed into 
a radiatively corrected, scale dependent light cone distribution function. 
It turns out that the relevant scales are small in both cases and 
different. The scale dependent distribution functions involve a 
convolution such that the nonperturbative effects cannot be cancelled 
anymore in the ratio.   

We have also computed the subleading terms partonically and use 
this to estimate the reliability of the leading twist comparison of 
$B \to X_s \gamma$ and $B \to X_u \ell \bar{\nu}_\ell$. This 
comparison depends on the width of the non-perturbative 
function once radiative corrections are taken into account. 
For values of $L_+$ close to the endpoint (i.e. within the width 
of the distribution function) one may use the leading twist 
contribution to extract $|V_{ub}| / |V_{ts}|$ with reduced hadronic 
uncertainties. This method should become feasible at the future 
$B$ physics experiments.

\section*{Acknowledgements}
We thank Kostja Chetyrkin, Timo van Ritbergen and Marek Jezabek for 
discussions and comments. SR and TM acknowledge the support of the DFG 
Graduiertenkolleg ``Elementarteilchenphysik an Beschleunigern''; 
TM acknowledges the support of the DFG Forschergruppe 
``Quantenfeldtheorie. Computeralgebra und MonteCarlo Simulationen''.

\end{document}